# Taylor cones of ionic liquids from capillary tubes as sources of pure ions: The role of surface tension and electrical conductivity

D. Garoz*, C. Bueno†, C. Larriba†, S. Castro†, Ignacio Romero-Sanz† and J. Fernandez de la Mora‡

*Yale University, Mechanical Engineering Department*

Y. Yoshida, G. Saito

*Division of Chemistry, Graduate School of Science, Kyoto University*

---

* Ph.D. Student, Actual address: E.T.S.I. Aeronáuticos, UPM.

Email: david.garoz@upm.es

† Ph.D. Student, Mechanical Engineering Dept., 9 Hillhouse Ave., New Haven, CT, 06520-8286

‡ Professor, Mechanical Engineering Dept., 9 Hillhouse Ave., New Haven, CT, 06520-8286

email: juan.delamora@yale.edu

## Abstract

The emissions of Taylor cones from a wide range of ionic liquids (ILs) have been tested in vacuo in an attempt to identify what physical properties favor the purely ionic regime (PIR). This regime is well known in the case of Taylor cones of liquid metals. For non-metallic liquids, it has been previously observed in conventional (capillary tube) electrospray sources at room temperature only for the room temperature molten salt (ionic liquid) EMI-$BF_4$ (EMI = 1-ethyl-3-methylimidazolium). A large number of other ILs and their mixtures have been studied here, most of which (but not all) are unable to reach PIR at room temperature. Based on these results and additional theoretical considerations, strong support is assembled for the notion that the PIR is favored not only by ionic liquids of high electrical conductivity, but also of high surface tension. This hypothesis is confirmed by tests with three new ILs EMI-$GaCl_4$, EMI-$C(CN)_3$, and EMI-$N(CN)_2$, all of which combine exceptional surface tension and electrical conductivity, and all of which reach the PIR at room temperature far more readily than EMI-$BF_4$.

## I. Introduction

When the meniscus of a conducting liquid is charged to a high voltage with respect to surrounding electrodes, it becomes conical and its sharp tip emits charged particles.[1,2,3,4,5] For liquids having electrical conductivities substantially below 1 S/m, the cone apex typically ejects a jet that breaks into a spray of charged drops. The drops have diameters of a few microns in the case of de-ionized water or other liquids having electrical conductivities in the range of $10^{-4}$ S/m. For this reason the phenomenon is commonly named an *electrospray*, though one should distinguish between the spray and the meniscus or *Taylor cone*[4]. Cloupeau and Prunet-Foch[5] have introduced the more specific term *cone-jet* to refer to the special regime of greatest practical interest, where a steady jet issues from the cone apex. In the case of liquids with exceptionally high electrical conductivities such as molten metals held in a vacuum, regimes exist where the emissions are essentially in the form of metal ions, so that the conventional long and unstable jet is replaced by a short stable tip from which metal ions are field evaporated. Such systems are referred to as liquid metal ion sources (LMIS), are in wide use in applications such as writing and etching on electrical circuits,[6] and have been proposed also as ion sources for electrical propulsion.[7] In between ion-emitting metals having electrical conductivities K larger than $10^{5}$ S/m and drop-producing dilute electrolytes with K values below $10^{-3}$ S/m, a transition from one behavior to the other is to be expected. It does indeed arise at conductivities in the range of 1 S/m,[8] but its investigation has been hampered by several difficulties. One is that, under atmospheric conditions, the ions generated in the presence of high electric fields attain ionizing energies, leading to electrical discharges that disrupt the Taylor cone. This problem may be overcome by operating under a vacuum. But a second difficulty then arises due to the lack of good solvents with the low volatility required to withstand the vacuum. Some polar but viscous solvents such as glycerol have been widely used as sources of drops and ions[9] in a vacuum, but they can barely reach K values approaching 0.02 S/m, which are insufficient to produce high quality ion beams. Only two neutral solvents capable of withstanding vacuum and forming electrolytes reaching conductivities of 1 S/m have been identified to date: formamide[8] and propylene carbonate (PC).[10.] Their Taylor cones in vacuum may produce a mixture of ions and drops, where the current of ions may be up to ten times higher than the drop current. These electrolytes, however are unsatisfactory because their finite volatility limits their practical value as electrical propellants in space applications, and precludes also well controlled terrestrial experiments. The presence of a finite fraction of drops mixed with the ions also leads to poor performance in electrical propulsion devices and other ion sources.

In view of these difficulties, and encouraged by the promising results reported long ago with $H_2SO_4$,[11] we have been studying the emissions from Taylor cones of similar (though less corrosive) purely ionic materials. Room temperature ionic liquids offer obvious advantages over salts melting at much higher temperature. Aside from the convenience of permitting room temperature operation, they tend to be far less reactive than inorganic molten salts. Furthermore, since the component ions of ionic liquids, especially the cations, are generally organic, they contribute a much wider range of chemical compositions and ion masses than conventional inorganic salts. For these reason, and also because most ionic liquids are essentially involatile and a fair number among them are known with room temperature electrical conductivities in excess of 1 S/m, we undertook their study[12] prior to that of high temperature molten salts. A first exploration[13] showed that the ionic liquid 1-ethyl-3-methylimidazolium bis(trifluoromethanesulfonyl)imide (EMI-$Tf_2N$) produced a mixture of ions and drops at comparable currents (mixed regime). However tests of the more conductive ionic liquid EMI-$BF_4$ revealed for the first time the existence of the sought purely ionic regime (PIR).[14] Subsequent investigations have confirmed this point with greater generality, and further shown a narrow ion energy distribution centered very close to the voltage of the emitting needle.[15,16,17] However, attempts to extend this behavior of EMI-$BF_4$ to other ionic liquids have been disappointing for reasons that remained unclear originally. For instance, Table I includes the ionic liquids studied by Romero,[18,19] of which only EMI-$BF_4$ exhibited the purely ionic regime at room temperature.

All but the most viscous among these liquids, however, did exhibit the PIR at the temperatures indicated in Table I.[18, 19] Similarly, Table II collects physical properties of several other ionic liquids synthesized and characterized by Garoz.[25] They include liquids formed by combinations of various amines with formic acid ($R_nNH_{4-n}$-COOH; n = 1-3; R = Me and Et), some of which can be seen to have exceptionally high electrical conductivities. This choice of materials was inspired by the work of Angell and colleagues.[20] who have argued that the proton transfer (neutralization) reaction between the acid and the base is partially reversible, leading to the presence of free protons yielding unusually high electrical conductivities (from their Figure 2B, at 25$^o$C, we read conductivities K ~ 1.7; 2.2 and 5.6 S/m, respectively, for $EtNH_3$-COOH, $EtNH_3$-$NO_3$ and a 4:6 molar mixture of $MeNH_3$-$NO_3$ and $Me_2NH_2$-$NO_3$.[26] More recent measurements give K = 4.38 and 1.22 S/m for $MeNH_3$-COOH and $EtNH_3$-COOH at 25$^o$C.[21] But because the small fraction of acid and amine remaining in equilibrium is volatile, the ionic liquids themselves are somewhat volatile, and this poses limitations to their use in a vacuum. Garoz[25] also tested several of his formate-based ionic liquids as ion emitters, but found substantial drop currents for all, including the singularly conducting salts based on reactions of amines with formic acid.

This early work provided clear evidence that the electrical conductivity was not the only parameter distinguishing between purely ionic or mixed regimes at room temperature. This was evident from a variety of observations, including the fact that EMI-$Tf_2N$ and $Me_2NH_2$-COOH do not reach the PIR at temperatures at which their electrical conductivity is substantially larger than the room temperature conductivity of EMI-$BF_4$. However, this negative information failed to provide a clue on which were the key properties besides electrical conductivity favoring the purely ionic regime. In the present study we first provide some details of the early studies of Garoz,[25] which remain unpublished. We then report new studies involving a number of other substances, indicating that another key property enabling PIR operation is high surface tension. This main conclusion has been previously summarized without detail in a study aimed at identifying ionic liquids having both high surface tension and high electrical conductivity.[24]

## II. Experimental

### A. Setup

The experimental setup is very similar to that of Gamero and Hruby.[13] The Taylor cones of various ionic liquids were supported on stainless steel needles (50 μm ID from MicroGroup) inside a vacuum chamber. Facing the needle perpendicularly to its axis is an extractor electrode (held at a slightly negative voltage trough an electrometer), with a small hole facing the needle and coaxial with it, through which the emitted charged particles entered a larger vacuum chamber. Periodic interruption of the Taylor cone via a high voltage switch that grounded it permits analysis of the beam by time of flight mass spectrometry (TOF-MS). The typical I(t)

TOF curve shown in Figure 1 represents the current I received by a collector electrode as a function of the time t following beam interruption. The trace I(t) exhibits a flat region at small t, followed by a succession of steps, each associated to the different times of flight of the various particles forming the beam. Short flight times reveal the presence of ions, and long flight times that of drops. Some of the characteristic quantities required to interpret the TOF data are included in the corresponding figures, where L is the distance between the extractor and the collector electrode, and V is the needle voltage. The collector was grounded through an electrometer, and was preceded shortly upstream by a grid held at a slight negative voltage. The liquid flow rate and the various propulsive parameters of interest have been inferred by suitable integration of the I(t) curves under the assumption that the energy of the charge drops is equal to their charge q times the needle voltage.[13] The liquid is fed to the Taylor cone from an external polypropylene vial into which the other end of the emitting capillary tube is immersed. The flow rate of liquid pushed through the capillary is controlled by varying the pressure $\Delta P$ of gas in the polypropylene reservoir.

The vacuum chamber is cylindrical, with 14.8 cm ID and 17.8 cm length. It was typically at a pressure of $10^{-5}$ mbar. Ion losses in the extractor have been measured and controlled during the experiments. These losses are usually less than 10 nA. But they are high in some special cases, such as the ionic regime with $Me_2NH_2$-COOH.

**B. Chemicals and syntheses**

Amines, tetrafluoroboric acid ($HBF_4$) and trifluoromethanesulfonic acid (H-TfO) were purchased from Aldrich. Formic acid was from Fluka, and H-$Tf_2N$ was a gift from Rhodia. $C_{14}(C_6)_3P$ salts were gifts from Professor John Wilkes or from Cytec. All syntheses are based on the neutralization of an amine with an acid, which requires certain essential safety precautions described by Angell and colleagues.[26] Note in particular that the reaction is highly exothermic and must proceed at low temperatures and small volumes. All the amines and acids are volatile and toxic and must be handled under a fume hood. Because of their high volatility the amines were purchased in aqueous solution, and the final products had to be dried under vacuum.

From the various neutralization reactions studied, only those included in Table II[25] produced salts that are liquid at room temperature. All the salts from amines neutralized with $HBF_4$ and H-TfO acid are solid at room temperature, except for $Et_3NH$-TfO, which is a very viscous liquid. However, the solid salts obtained have low melting temperature, and some of them like $Et_3NH$-$BF_4$ are liquid at room temperature when mixed with a moderate percentage of other low melting point ionic liquids such as EMI-$Tf_2N$. Taylor cones of these mixtures were briefly studied, but did not give promising results. Finally two salts formed from amines neutralized with H-$Tf_2N$ are liquid at room temperature, and are included at the end of Table II. Although their conductivities below 1 S/m do not favor the ionic regimes, one of them was found to be very useful in mixtures (Section IV). $MeNH_3$-$Tf_2N$ and $Et_2NH_2$-$Tf_2N$ are both solid at room temperature. $EtNH_3$-$Tf_2N$ and $Me_3NH$-$Tf_2N$ are also solid, but melt relatively close to room temperature (51-84$^o$C for $Me_3NH$-$Tf_2N$, and about 36$^o$C for the $EtNH_3$-$Tf_2N$).

**C. Physical properties of the new salts**

The properties described in Table II[25] were obtained at room temperature. Densities, measured with a precision balance and a micropipette, take into account scale errors and use least square method with three measurements. Approximate electrical conductivities were determined by a method due to Professor John B. Fenn based on measuring the static electrical resistance of the liquid under study held inside a Teflon tube (0.31 mm ID). Errors from voltmeter scale, resistance and geometrical ambiguities have been considered in error estimates. Total errors have been minimized taking five measurements and using least square method.

**D. Volatility of the ionic liquids**

A qualitative measure of volatility is given in Table II, as determined by comparing the time required for size reduction of a drop of 1.5 µl placed over a Teflon base within a vacuum chamber held at 10 mbar. As reference we used the corresponding times for the organic liquids formamide and propylene carbonate, with boiling points 210ºC (decomposes) and 240ºC respectively. Propylene carbonate is the most volatile substance (*very high* volatility; time to disappear ~ 2 hours), then $Me_3NH$-COOH and $Et_3NH$-COOH are *highly* volatile (3.5 hours). Formamide has a *moderate* volatility (5.5 hours). $Me_2NH_2$-COOH (6.5 hours) has a *small* volatility. The rest, $MeNH_3$-COOH, $EtNH_3$-COOH and $Et_2NH_2$-COOH have even lower volatilities (11.5 hours). The salts with the $Tf_2N^-$ anion that remained liquid at room temperature ($Et_3NH$- $Tf_2N$ and $Me_2NH_2$-$Tf_2N$) are considerably less volatile than the formates, as they did not lose appreciable mass after several days of exposure to a vacuum. Other experiments at lower pressures gave comparable results, though with smaller evaporation times. For instance, with a background at 0.020 mbar, 2µl drops survive for 4 min (PC), 7min ($Et_3NH$-COOH), 19 min ($Met_3HN$-COOH), 35 min ($Met_2H_2N$-COOH), and 52 min (tributylphosphate). $MetH_3N$-COOH, $EtH_3N$-COOH and $Et_2H_2N$-COOH last 80 min, while water freezes and then sublimes in 15 min. In these experiments the evaporation time is counted from the moment when the mechanical pump is turned on, though it takes about 5 minutes for the pressure to drop from 0.1 mbar to 0.02 mbar. Other indications on volatility of ILs are given in Ref.[27]

## III. Electrospraying characteristics of the formate salts

Figure 1 shows TOF curves for Taylor cone emissions from several selected formates. The results for $Et_2NH_2$-COOH and $Et_3NH$-COOH are similar to those for $EtNH_3$-COOH (Figure 1d) and are not shown for brevity. Note first that none of the liquids achieves the purely ionic regime. Still, $MeNH_3$-COOH, $Me_2NH_2$-COOH and $Me_3NH$-COOH produce ions abundantly, while the emissions from ethylammonium formates are dominated by drops. Among the first group, note the much smaller tail of drops in the case of the singularly conducting $Me_2NH_2$-COOH. Figures 1a-c show the ion-dominated regimes of $MeNH_3$-COOH, $Me_2NH_2$-COOH and $Me_3NH$-COOH, with increasing currents associated to increasing needle voltage V or liquid reservoir pressure $\Delta P$. $Me_3NH$-COOH produces a large drop current under most conditions, but it yields high ionic currents at increased voltages (Figure 1c). Using wider needles with $MeNH_3$-COOH, one finds large drop currents at low voltages and pressures, and high ionic currents at increased voltages or pressures, similarly as with $Me_3NH$-COOH. Accordingly, relatively small changes of conditions vary greatly the specific impulse. The ion-dominated regime of the formates yields very high currents, with relatively large thrust and specific impulse. However, the propulsion efficiency is low due to the mixed emissions of ions and drops. Figure 2a shows several propulsion parameters exhibiting two different regions. At currents above 7 $\mu$A the emissions from $Me_2NH_2$-COOH are dominated by ions, with relatively high specific impulses and significant ion losses in the extractor. The latter are due to the fact that that the solid angle occupied by the spray increases rapidly with total current. For currents below 7 $\mu$A, $MeNH_3$-COOH, $Me_2NH_2$-COOH and $Me_3NH$-COOH have similar propulsion parameters with low efficiencies (Table III). Due to the electrochemical formation of gas bubbles that destabilize the Taylor cone, $MeNH_3$-COOH and $Me_3NH$-COOH do not achieve the ion-dominated regime under negative polarity. Neither the drop-dominated regime (both polarities), nor the ion-dominated conditions in positive mode present this difficulty. $Me_2NH_2$-COOH can be electrosprayed in both modes, positive and negative, yielding similar propulsion parameters in both polarities.

The emissions from $EtNH_3$-COOH, $Et_2NH_2$-COOH and $Et_3NH$-COOH are dominated by drops, though they include also small ion currents. All three can be electrosprayed in both polarities, as shown in Figure 1d. At the acceleration voltage used ~2000 V the best specific impulse (Isp) obtained with $EtNH_3$-COOH is below 200 s (Table III), worse than attainable by chemical propellants. However, post acceleration to 10 kV would yield a more respectable Isp ~ 450 s. Single Taylor cone thrust levels for these three liquids are in the range of 1µN, with an approximate linear dependence on total current (Figure 2b). Propulsion efficiencies are reasonably high ~ 0.8, but the small ion currents emitted preclude improving this parameter further. It is possible to attain ion-dominated regimes with $EtNH_3$-COOH and $Et_3NH$-COOH by using needles of wider diameter together with high voltages. But the TOF traces exhibit relatively long tails of large drops leading to poor propulsive characteristics.

## IV. Propellants based on ionic liquids combining amines and the $Tf_2N^-$ anion

In view of the volatility of the formate salts just discussed, we have also tested Garoz's two room temperature ionic liquids including the $Tf_2N^-$ anion. Unfortunately, as shown in Figure 3 for $Et_3NH$-$Tf_2N$, these salts have insufficient conductivity to achieve the purely ionic regime.

Even though many $Tf_2N^-$ based salts are highly hydrophobic, $Et_3NH$-$Tf_2N$ and $Me_2NH_2$-$Tf_2N$ are miscible with a number of substances, including hydrophobic and hydrophilic ionic liquids. For instance, $Et_3NH$-$Tf_2N$ and $Me_2NH_2$-$Tf_2N$ both mix at 50% (v/v) with $MeNH_3$-

COOH and with EMI-$Tf_2N$. Even more interesting, although $Me_2NH_2$-$Tf_2N$ is not very miscible with EMI-$BF_4$, $Et_3NH$-$Tf_2N$ is miscible with EMI-$BF_4$ at all concentrations, with an electrical conductivity varying monotonically from one end to the other (Table IV). This property is of considerable interest, as it enables a continuous transition from propellants running in the purely ionic regime (EMI-$BF_4$ rich mixtures) to others running only on the mixed regime (ions + drops; $Et_3NH$-$Tf_2N$ rich mixtures). Indeed, as shown also in the bottom row of Table IV, the Taylor cones from these mixtures exhibit a sharp transition at a critical electrical conductivity of about 0.72 S/m, below which the emissions are mixed, and above which they are purely ionic. It is striking that either one behavior or the other arises almost discontinuously as physical properties vary continuously in these ionic liquid mixtures. Also noteworthy is the relatively small electrical conductivity at which the transition occurs. Note the discrepancy between the conductivity of the pure $Et_3NH$-$Tf_2N$ in Tables II and IV. The conductivity data in Table IV are more reliable, as they were all taken with the same Teflon tube, calibrated based on the published conductivity of pure EMI-$BF_4$.

## V. EMI-TfO almost achieves the purely ionic regime

The ionic liquid EMI-TfO is commercially available and commonly used in conventional ionic liquid applications. Its electrical conductivity is comparable to that of EMI-$Tf_2N$, and because EMI-$Tf_2N$ does not exhibit the purely ionic regime, EMI-TfO was presumed to be similarly limited. However, while attempting to understand the reasons why certain pure or mixed ionic liquids exhibit the purely ionic regime and others do not, we noted that the surface tension of EMI-$BF_4$ is considerably higher than that of other liquids. In order to determine if this singularity could in part explain the superior behavior of EMI-$BF_4$, we examined preliminary surface tension measurements reported by Romero[18] and noted that EMI-TfO had the highest surface tension following EMI-$BF_4$. This realization suggested studying the emissions from its Taylor cones, with results shown in Figure 5. Interestingly, this liquid operates almost at the purely ionic regime.

## VI. The role of surface tension

The notion that a high surface tension plays a large role in promoting the purely ionic regime is in fact confirmed by all the liquids previously discussed. For instance, Garoz[25] had measured the surface tension of $EtNH_3$-COOH and obtained at room temperature the relatively high value $\gamma$ = 43.8 dyn $cm^{-1}$. Yet according to Martino et al.[22], atmospheric humidity rapidly contaminates the surface of vacuum-dried hydrophilic liquids, increasing drastically their surface tension. It is therefore essential for these materials to keep the surface dry during the measurement, a precaution not taken in the early measurements of either Garoz[25] or Romero.[18] Our new measurement for the surface tension of $EtNH_3$-COOH under a dry atmosphere gives the smaller value $\gamma$ = 37.35 dyn $cm^{-1}$ at 23 $^oC$.[24] We have similarly measured the surface tension of $Et_3NH$-$Tf_2N$, and found it to be even smaller ($\gamma$ = 30.19 dyn $cm^{-1}$ at 23 $^oC$).[24] Accordingly, when reinterpreted with correct surface tension data, these early studies strongly suggest the proposition that both high surface tension and high electrical conductivity favor PIR operation. This hypothesis is also in accord with what one would expect on purely theoretical grounds. The scaling laws for the structure of cone-jets indicate that the maximum electric field on the surface of the meniscus varies as[8]

$$E_{max} = \varphi(\varepsilon)\, \gamma^{1/2} \varepsilon_o^{-2/3} (K/Q)^{1/6}, \qquad (1)$$

where $\varepsilon_o$ is the electrical permittivity of vacuum, and Q is the flow rate of conducting liquid pushed through the meniscus tip and the jet. $\varphi(\varepsilon)$ is a proportionality coefficient of order unity that depends on the dielectric constant $\varepsilon$. This law (1) has been recently confirmed numerically[10] with a value of $\varphi = 0.76$ for propylene carbonate ($\varepsilon = 65$). Q cannot be arbitrarily small due to the finite range of stability of Taylor cones. The minimal value it can take scales as $\gamma/K$,[23] whence $E_{max} \sim (\gamma K)^{1/3}$, confirming the role expected of both $\gamma$ and K to attain high surface electric fields. The importance of the surface electric field itself follows of course from its role in reducing the activation barrier for an ion to evaporate. Because this barrier is typically 1.8 eV, in the absence of intense electric fields, ion evaporation rates are astronomically small at room temperature.[24]

## VII. Mixtures of ionic liquids with acids.

The possibility to vary continuously the physical properties of the liquid and observe its effects on the resulting electrosprays is of great interest, as evident from our findings with mixtures of $Et_3NH$-$Tf_2N$ + EMI-$BF_4$. One of the related strategies explored has been based on the addition of acids to various ionic liquids, with the purpose of enhancing their electrical conductivity. The rationale for this approach was an expectation that most properties of the ionic liquid would remain fixed, while the electrical conductivity would change greatly with slight acidification. In order to avoid complications associated to displacement reactions between the acid and the salt, the initial (and most promising) studies restricted the mixtures to salts and acids sharing the same anion. Due to the volatility of most acids involved in the previously discussed salts, such mixtures would appear in principle as unsuitable for vacuum work. However, during the ionic liquid synthesis process we had noted that, when the products were slightly acidic, it was not possible to make them neutral by keeping the liquid under vacuum for many hours. It was therefore clear that acid evaporation from the melt was strongly opposed by its interaction with the salt. We shall later comment on the weaknesses of our thinking on the effects of acid addition, and start by reporting the results obtained.

Table V summarizes the mixtures investigated, including their electrical conductivities at room temperature and whether or not corresponding TOF spectra were taken. No surface tensions have been measured for these mixtures.

The only mixtures where a substantial advantage was observed involved EMI-$Tf_2N$ + H-$Tf_2N$. As shown in Figure 6, a few percent addition of acid led to emissions close to the purely ionic regime. The improvement in performance is striking, as neat EMI-$Tf_2N$ sprays from capillary needles typically include drop currents of 50% or more.[13]

We have also examined acidic mixtures including two different anions. Those of EMI-$BF_4$ and H-$Tf_2N$ become less conducting than the pure ionic liquid. It would have perhaps been more interesting to study mixtures EMI-$BF_4$ + $HBF_4$, but this was not done. The mixtures of EMI-TfO and H-$Tf_2N$ showed a modest increase in electrical conductivity, and a slight tendency to reduce the drop contribution from the already almost purely ionic emissions from neat EMI-TfO. The other combinations with heterogeneous anions shown at the bottom of Table V did not appear as sufficiently conducting to be pursued.

Recent work[25] has begun to clarify the mechanisms involved in the reduced volatility of non-stoichiometric melts of acids (or bases) and salts. For instance, in the system EMI-Cl/HCl, excess protons are not free, but are instead sequestered in the complex $HCl_2^-$, precluding therefore acid evaporation. Notice as a corollary that acid addition does not necessarily contribute high mobility species, but may in fact reduce the electrical conductivity of an ionic liquid. In contrast, in the case of acid/base neutralization, base-rich melts can exhibit enhanced conductivity via the Grotthuss mechanism, where the excess base (say amine) is available for proton transfer from a protonated amine, providing a singularly efficient proton jumping transport alternative to conventional conduction. Unfortunately this knowledge has not yet been exploited for electrical propulsion applications.

## VIII. Emissions from ILs of high surface tension and electrical conductivity

The many observations so far reported suggest that PIR operation requires liquids with high surface tension and high electrical conductivity. As a result, we have embarked in a broad effort to identify ionic liquids besides EMI-$BF_4$ and EMI-TfO having such desired characteristics. A large number of them have been found,[28,24] though the systematic investigation of the composition of their electrosprays is still underway. We can nonetheless advance here a certain level of confirmation of the proposition that the combination of high surface tension and high electrical conductivity produces outstanding purely ionic emitters. Figure 7 shows TOF spectra for EMI-$BF_4$ together with those for three other liquids having exceptionally high electrical conductivity and surface tension (Table VI), even by the already rather high standards of EMI-$BF_4$ (the previous value[28] $\gamma = 42.6$ dyn cm$^{-1}$ for EMI-$N(CN)_2$ was incorrect). Also included in the last row are data for the newly synthesized EMP-$N(CN)_2$ (EMP: 1-ethyl-2-methylpyrazolium),[34] which has a singular surface tension, but a smaller electrical conductivity than the other three new ILs. The emitters are in this case silica capillaries 360 μm in outer diameter, sharpened conically to a tip diameter relatively close to their inner diameter of 40 μm. The needle itself is not conducting, so the high voltage electrode is introduced at the liquid reservoir, and the charge reaches the cone by conduction through the liquid inside the capillary. For the 40 μm silica capillaries used here, this leads typically to a 20% ohmic voltage drop through the line.

As seen in Figure 7, the three new liquids $A^+B^-$ with both high K and γ achieve cleanly the PIR, with only the monomer $A^+$ and dimer $(A^+)_2B^-$ ions (first and second steps) present. In contrast, EMI-$BF_4$ exhibits a tail extending to about 70 μs, corresponding to a highest mass/charge ratio of some 11,000 Dalton (100 times larger than that of the $EMI^+$ ion). Almost complete suppression of this tail is possible in EMI-$BF_4$ sprays, but it is sometimes difficult for reasons that remain unclear. Similar or greater difficulties arise in the case of the other inferior ILs for which we have already noted an ability to approach purely ionic emissions. In contrast, the three new ILs, EMI-$GaCl_4$, EMI-$C(CN)_3$, and EMI-$N(CN)_2$, are considerably more favorable in this respect. Although they can also be forced to eject drops at larger liquid flow rates, all have a wide range of readily achieved operational conditions where only ions are produced.

The case of EMP-$N(CN)_2$ is of interest as it is intermediate between the two other emission patterns just discussed. There is a high mass tail, but it extends only to about 2,500 Dalton. It is noteworthy that the transition leading to the appearance of these high mass tails seems to be correlated with a decrease in electrical conductivity (or an increase in viscosity), and that the singular surface tension of EMP-$N(CN)_2$ does not mitigate the negative effects of the reduced electrical conductivity.

In conclusion, the best results have been obtained with ILs with γ approaching 50 dyn $cm^{-1}$ and K exceeding 2 S/m. Borderline behaviour is observed at K ≤1.7 S/m [EMP-$N(CN)_2$] even for higher surface tension ILs. Mixed emissions are observed at conductivities above 5 S/m [$Me_2NH_2$-COOH] at a surface tension not yet determined, but surely considerably lower than 50 dyn cm$^{-1}$. This conclusion on the desirability for both high $\gamma$ and high K, however, should be balanced by noting recent work showing that a different source type referred to as ILIS, based not on a capillary tube but on an externally wetted electrochemically roughened and sharpened tungsten needle,[15] achieves purely ionic emission of EMI-$Tf_2N$ (K = 0.88; $\gamma$ = 34.9 dyn cm$^{-1}$).[26] A companion paper[27] on such ionic liquid ion source (ILIS) extends to a wider range of ILs Lozano's discovery[36] that the new source is able to produce purely ionic beams from some liquids with relatively modest values of both the surface tension and the electrical conductivity. Particularly noteworthy is the fact that the largest clusters formed by EMP-$N(CN)_2$ have sizes comparable to those recently reported for far less conducting ILs emitted from ILIS sources.[28] These clusters are much smaller than the smallest drops previously observed in pure drop emission from Taylor cones, yet much larger than can be explained by ion evaporation theory.[38] There is hence no current explanation for their appearance in either of these two quite different ion sources.

## Acknowledgment

We are thankful to Prof. Martinez Sanchez (MIT) for many contributions to the electrospraying aspects of the problem, and to Professor J. Wilkes (US Air Force Academy) and Dr. A. Robertson (Cytec) for their kind gift of several ionic liquid samples. Our work has been supported by the US AFOSR through grants F-49620-01-1-0416 and FA9550-06-1-0104 at Yale, and Yale subcontracts to STTR grants to the companies Busek and Connecticut Analytical Corporation.

Table I. Ionic liquids studied in[18,19], including key physical parameters and the temperature $T_{PIR}$ at which their Taylor cones achieve the purely ionic regime.

| Liquid | $\gamma$ (dyn cm$^{-1}$) | $\rho$ (g cm$^{-3}$) | $\mu$ (cP) | K (S/m) | Reference | $T_{PIR}$ (°C) |
|---|---|---|---|---|---|---|
| EMI-$BF_4$ | 45.2 (23°C)[d] | 1.24 | 38 | 1.4 | [29] | < 23 |
| EMI-$Tf_2N$ | 34.9[a] | 1.52 | 34 | 0.88 | [20] | < 82 |
| EMI-$(C_2F_5SO_2)_2N$ | 28.75(21°C)[d] | 1.6[a] | 61 | 0.34 | [30] | < 119 |
| EMI-$(CF_3SO_2)_3C$ | 32.4[a] | 1.496[a] | 195[b] | 0.13 | [20] | < 216 |
| $C_4$MI-$(C_2F_5SO_2)_2N$ | 27.6[a] | 1.425[a] | | | | < 204 |
| DMPI-$(C_2F_5SO_2)_2N$ | 29.7[a] | 1.506[a] | | | [20] | < 212 |
| DMPI-$(CF_3SO_2)_3C$ | 37.8[a] | 1.55[a] | 726[b] | 0.046 | [20,31] | None |
| $C_{14}(C_6)_3$P-$Tf_2N$ | 27.4[a] | 1.021[a] | | | | None |
| $C_{14}(C_6)_3$P-$N(CN)_2$ | 29.2[a] | 0.987[a] | | | | None |
| $C_5$MI-$PF_3(C_2F_5)_3$[c] | 28.75 (21°C)[d] | 1.59 | 88.4 | 0.166 | [32] | < 157 |

<u>Cations:</u> EMI = 1-ethyl-3-methylimidazolium; DMPI = 1,2-dimethyl-3-propylimidazolium; $C_4$MI = 1-butyl-3-methylimidazolium; $C_5$MI = 1-pentyl-3-methylimidazolium; $C_{14}(C_6)_3$P = trihexyltetradecylphosphonium, commercialyzed by Cytec under the name Cyphos.

<u>Anions:</u> $Tf_2N = (CF_3SO_2)_2N$.

[a] Rough measurements without drying precautions.

[b] Private communication, V. Koch, Covalent.

[c] Merck home page.

[d] From Table IV of[33]

Table II. Some characteristics of the ionic liquids synthesized by Garoz.[34]

| Liquid | FW (amu) | ρ (g cm$^{-3}$) | γ (dyn cm$^{-1}$) | K (S/m) | Volatility |
|---|---|---|---|---|---|
| $MeNH_3$-COOH | 77.09 | 1.12 | | 2.9±0.3 | Low |
| $Me_2NH_2$-COOH | 91.11 | 1.06 | | 6.7±0.9 | Moderate |
| $Me_3NH$-COOH | 105.14 | 1.09 | | 3.8±0.8 | High |
| $EtNH_3$-COOH | 91.11 | 1.1 | 43.8[a]; 37.35[b] | 1.4±0.1 | Low |
| $Et_2NH_2$-COOH | 119.17 | 1.02 | | 0.96±0.09 | Low |
| $Et_3NH$-COOH | 147.23 | 1.06 | | 1.04±0.09 | High |
| $(C_8H_{17})_3NH$-COOH | 399.71 | 0.84 | | 0.014±0.001 | - |
| $Et_3NH$-TfO | 251.3 | 1.19 | | 0.46±0.06 | - |
| $Me_2NH_2$-$Tf_2N$ | 326.23 | 1.48 | | 0.61±0.04 | - |
| $Et_3NH$-$Tf_2N$ | 382.35 | 1.41 | 30.19[b] | 0.52±0.06 | - |

Anions: TfO = $CF_3SO_3$

[a] Under humid atmosphere. Note that the water content was not measured, while most salts contained a slight excess of base.

[b] From Larriba et al.[24]

Table III. Propulsive characteristics of several formate propellants

| Ionic liquid | Voltage (volts) | Pressure (Torr) | $I_{total}$ (nA) | $I_{extractor}$ (nA) | Thrust (μN) | Flow rate (μg $s^{-1}$) | Isp (s) | Propulsion efficiency |
|---|---|---|---|---|---|---|---|---|
| $MeNH_3$-COOH | 1980 | 15 | 1430 | -1 | 0.381 | 0.114 | 340 | 0.25 |
| | 2060 | 40 | 2220 | -4 | 0.406 | 0.076 | **554** | **0.26** |
| | 2000 | 100 | **4520** | 2 | **0.778** | 0.181 | 439 | 0.21 |
| $Me_2NH_2$-COOH | 2346 | 1 | 2260 | 5 | 0.402 | 0.0597 | **687** | 0.28 |
| | 1898 | 11.5 | 2660 | 600 | 0.446 | 0.109 | 419 | 0.2 |
| | 1763 | 21.9 | 5360 | 160 | 0.892 | 0.275 | 331 | 0.17 |
| | 2061 | 45 | 8000 | 1600 | 0.843 | 0.0755 | 1140 | 0.32 |
| | 2142 | 88.9 | **14500** | 2400 | **1.46** | 0.128 | **1168** | 0.3 |
| $Me_3NH$-COOH | 2260 | 47 | 720 | 0 | 0.578 | 0.245 | **241** | 0.46 |
| | 2000 | 122 | **930** | 127 | **1.33** | 1.01 | 139 | 0.52 |
| | -1920 | 160 | **1080** | -67 | **1.74** | 1.43 | 124 | 0.57 |
| $EtNH_3$-COOH | 1960 | 197 | 340 | 0 | 0.515 | 0.296 | **178** | 0.75 |
| | 2180 | 535 | **600** | 0 | **1.44** | 1.24 | 118 | 0.7 |
| | -1820 | 122.4 | 456 | 0 | 0.608 | 0.353 | **176** | 0.71 |
| | -2040 | 310.0 | **740** | -10 | **1.5** | 1.2 | 127 | 0.68 |
| $Et_2NH_2$-COOH | 1530 | 85 | 228 | 0 | 0.241 | 0.118 | 209 | 0.82 |
| | 1720 | 85 | 220 | 0 | 0.261 | 0.119 | **224** | **0.86** |
| | 1640 | 310 | 472 | 1 | 1.09 | 1.12 | 100 | 0.78 |
| | 2200 | 310 | **544** | 0 | **1.35** | 1.1 | 126 | 0.76 |
| $Et_3NH$-COOH | 2160 | 85 | 344 | 1 | 0.627 | 0.365 | **175** | 0.8 |
| | 1620 | 160 | 416 | 1 | 0.894 | 0.802 | 114 | **0.84** |
| | 2280 | 197 | **632** | 0 | **1.68** | 1.56 | 110 | 0.69 |

Table IV. Characteristics of mixtures of $Et_3NH$-$Tf_2N$ + EMI-$BF_4$ as a function of EMI-$BF_4$ concentration.

| Vol % EMI-$BF_4$ | 0 | 10 | 30 | 50 | 60 | 65 | 70 | 80 | 100 |
|---|---|---|---|---|---|---|---|---|---|
| K(S/m) | 0.39 | 0.53 | 0.65 | 0.66 | 0.74 | 0.72 | 0.82 | 0.99 | 1.3 |
| Regime | M | M | M | M | M | M&I | I | I | I |

M = mixed regime; I = purely ionic regime; M&I means that both regimes are attainable at high and low voltages, respectively.

Table V. Electrical conductivities of several ionic liquids seeded with H-$Tf_2N$

| | K (S/m) | TOF |
|---|---|---|
| EMI-$Tf_2N$ + H-$Tf_2N$ 1% | | Yes |
| EMI-$Tf_2N$ + H-$Tf_2N$ 2 % | 0.85 | Yes |
| EMI-$Tf_2N$ + H-$Tf_2N$ 5 % | 0.87 | Yes |
| EMI-$BF_4$ + H-$Tf_2N$ 1% | 1.30 | Yes |
| EMI-$BF_4$ + H-$Tf_2N$ 5% | 1.34 | Yes |
| EMI-TfO | 0.87 | Yes |
| EMI-TfO + H-$Tf_2N$ 2% | 0.91 | Yes |
| EMI-TfO + H-$Tf_2N$ 5% | 0.97 | Yes |
| EMI-$(C_2F_5SO_2)_2N$ + H-$Tf_2N$ 2% | 0.47 | No |
| $C_5$MI-$(C_2F_5)_3PF_3$ + H-$Tf_2N$ 2% | 0.37 | No |

Table VI. Physical properties of the ionic liquids used in Figure 7.

| Liquid | γ(at 23ºC) (dyn $cm^{-1}$) | ρ (g $cm^{-3}$) | K (S/m) | η (cP) |
|---|---|---|---|---|
| EMI-$BF_4$ [24][a] | 45.2[b] | 1.24 | 1.4 | 38 |
| EMI-$GaCl_4$ **[**35**]**[a] | 48.6 (21ºC)[b] | 1.53 | 2.2 | 16 |
| EMI-$C(CN)_3$ [36][a] | 47.9[b] | 1.11 | 2.2 | 18 |
| EMI-$N(CN)_2$[c] [33][a] | 49.05[d] | 1.08 | 2.8 | 16 |
| EMP-$N(CN)_2$ [37][a] | 60.69[d] | 1.16 | 1.7 | 25 |

Cations: EMP = 1-ethyl-2-methylpyrazolium

[a] Values for ρ, K and η are from the reference indicated in the first column.

[b] Reference [24]

[c] For the original synthesis see[38]

[d] New data from the present work.

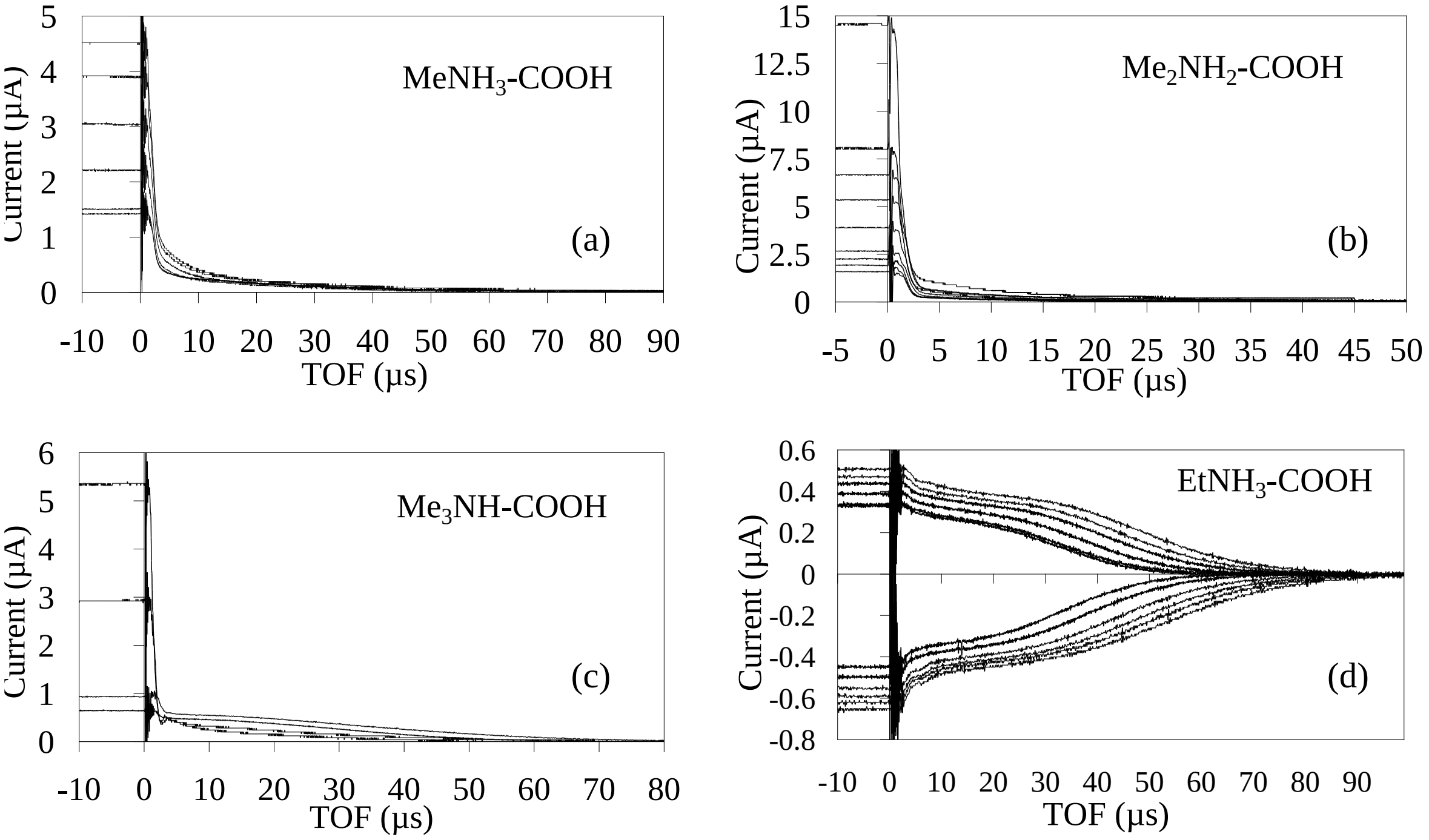

Figure 1. TOF curves of formates, using a stainless steel needle (ID 50 µm). The highest current corresponds to the highest voltage V and pressures ΔP on the liquid reservoir. (a) $MeNH_3$-COOH; V = 2000 volts ; L = 10.6 cm; ΔP = 15 Torr to 100 Torr. (b) $Me_2NH_2$-COOH; L = 7.3 cm; varying V and ΔP. (c) $Me_3NH$-COOH; L = 6.3 cm; ΔP = 122 Torr; V from 1900 to 2400 volts; over 2100 volts the regimens are ionic. (d) $EtNH_3$-COOH represented positive (1800 volts) and negative (1770 volts); L = 6.3 cm; ΔP from 160 to 500 Torr.

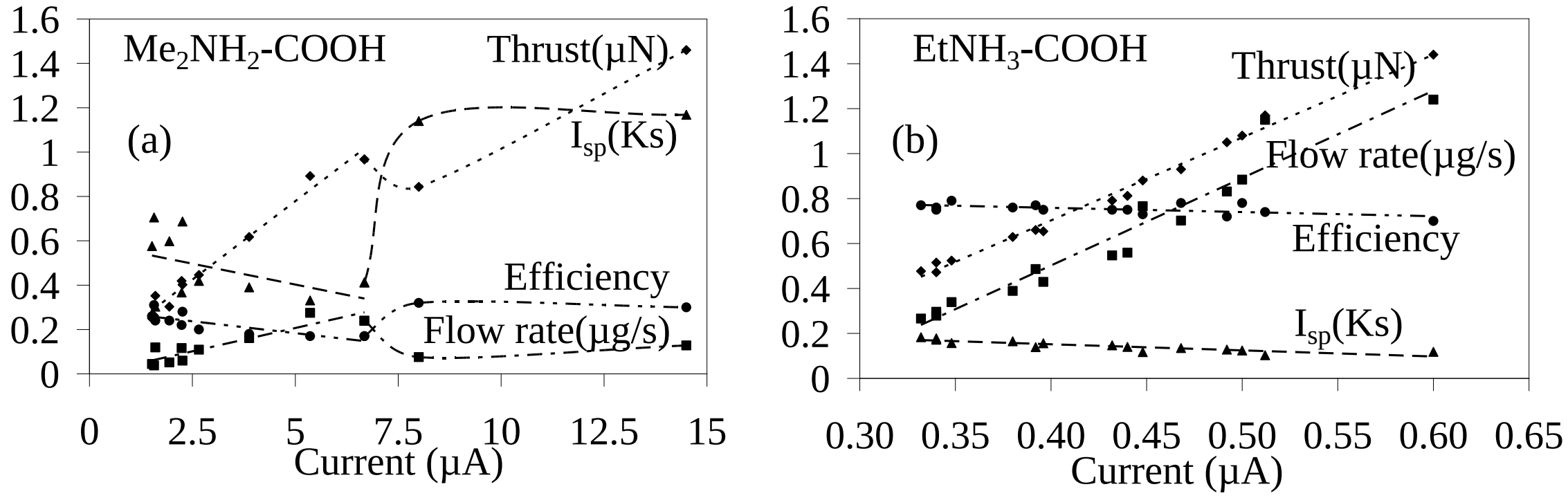

Figure 2. Representative propulsion characteristics of the formate-based ionic liquids, with units indicated on each curve. For $Me_2NH_2$-COOH (a), note that high currents losses take place in the extractor. $EtNH_3$-COOH (b), runs in the drop-dominated regime with low Isp and fair propulsion efficiency.

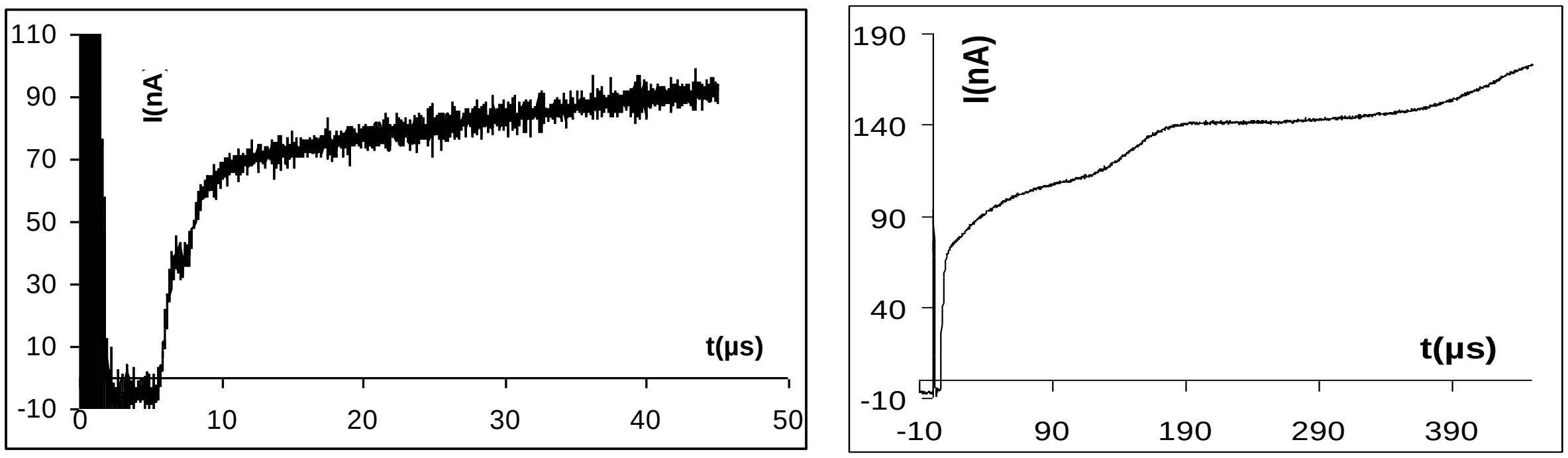

Figure 3: TOF spectra for electrosprays of pure $Et_3NH$-$Tf_2N$. The left Figure is a detail at shorter times, showing the steps associated to the ions. The Figure to the right includes the full spectrum showing a large tail of charged drops. V = 1532 volts; $\Delta P$ = 15 Torr; needle ID 150 μm.

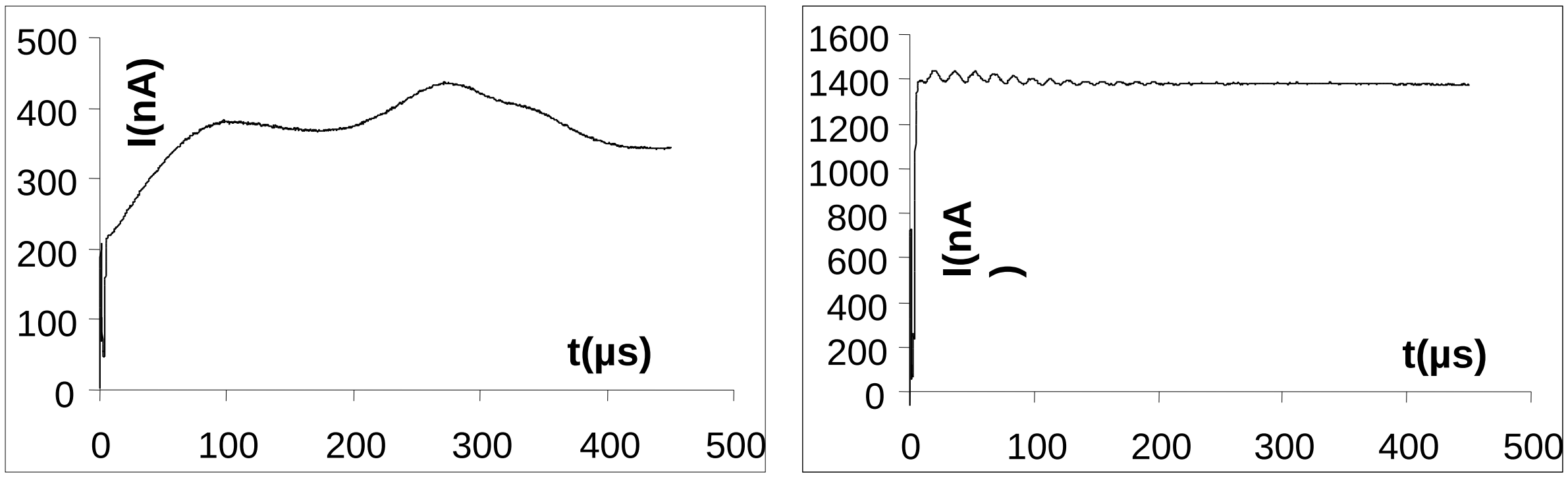

Figure 4: Emissions from critical mixture of $Et_3NH$-$Tf_2N$ + $EMIBF_4$ (35-65 % vol.), below which the purely ionic regime (right) is sometimes observed. The mixed regime may also arise (left; V = 1883 volts, $V_{extractor}$ = -700 volts, ΔP = 10 Torr). Although the purely ionic regime arises at higher voltages, it remains “put” as the voltage is later reduced (right; V = 1829 volts, $V_{extractor}$ = -700 volts, ΔP = 10 Torr).

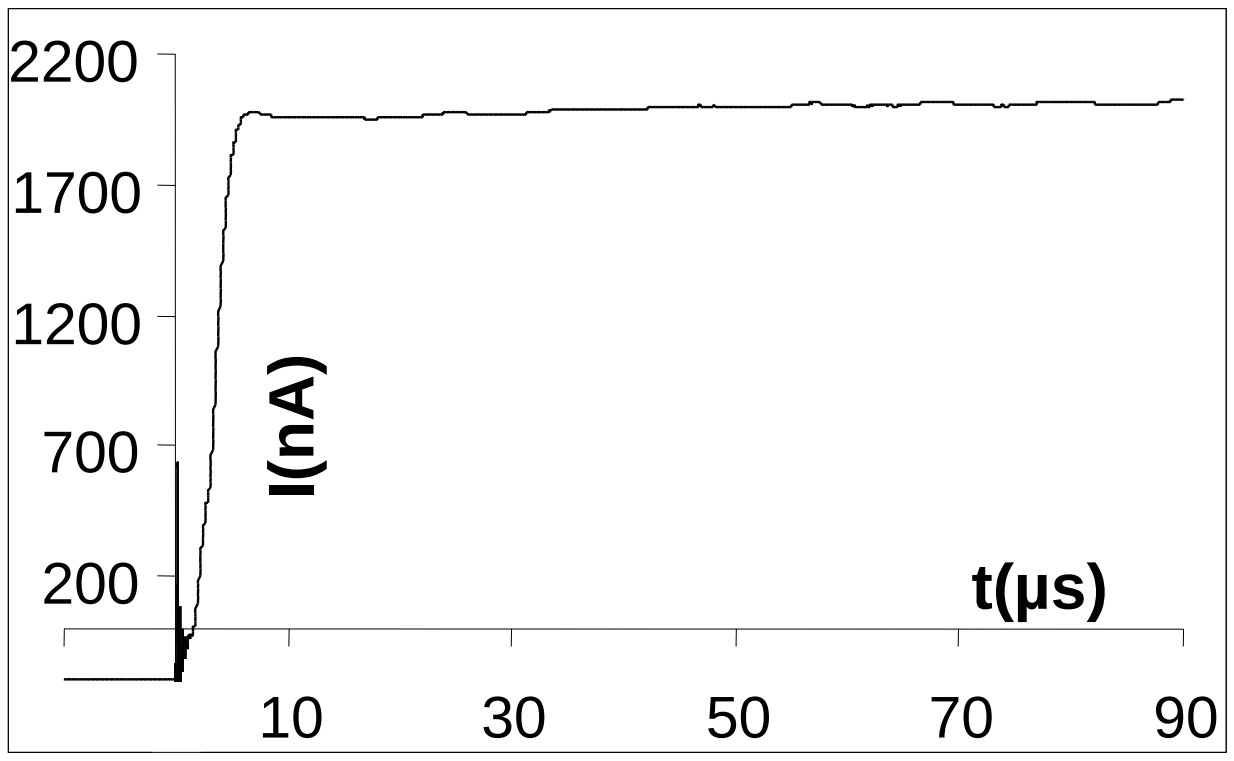

Figure 5: TOF spectrum for an electrospray of pure EMI-TfO, showing an essentially purely ionic emission. ΔP =10 Torr; V=2440 volts.

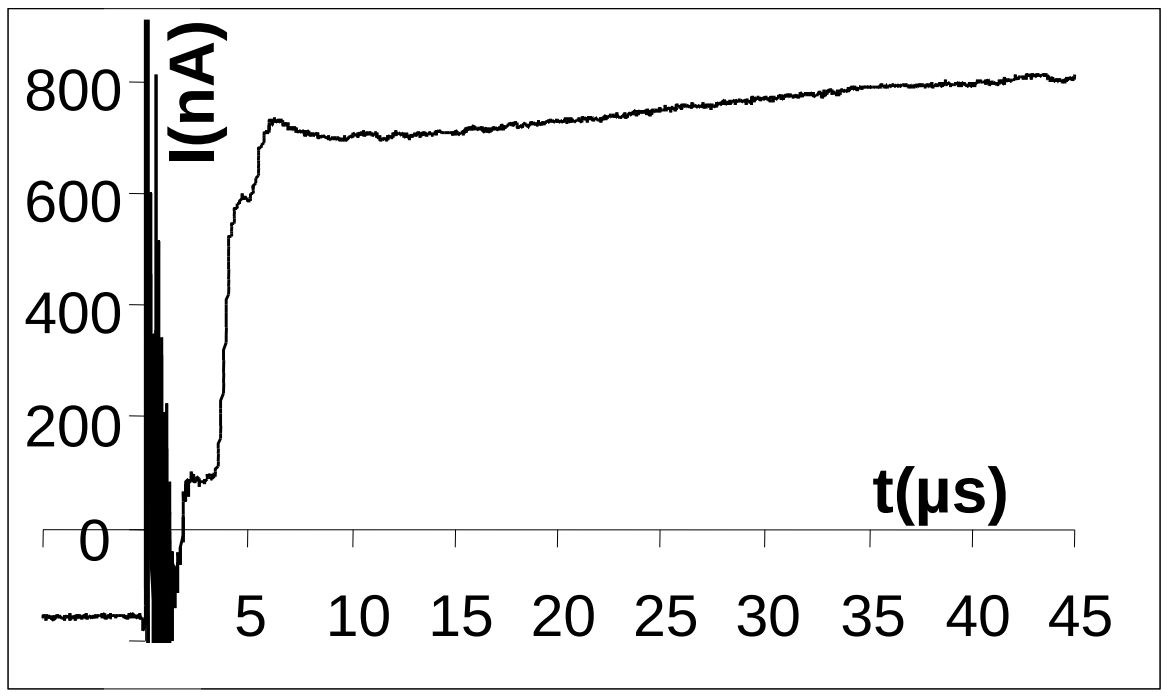

Figure 6: TOF spectrum of EMI-$Tf_2N$ seeded with H-$Tf_2N$ acid (1%v/v), showing a behavior close to the purely ionic regime.

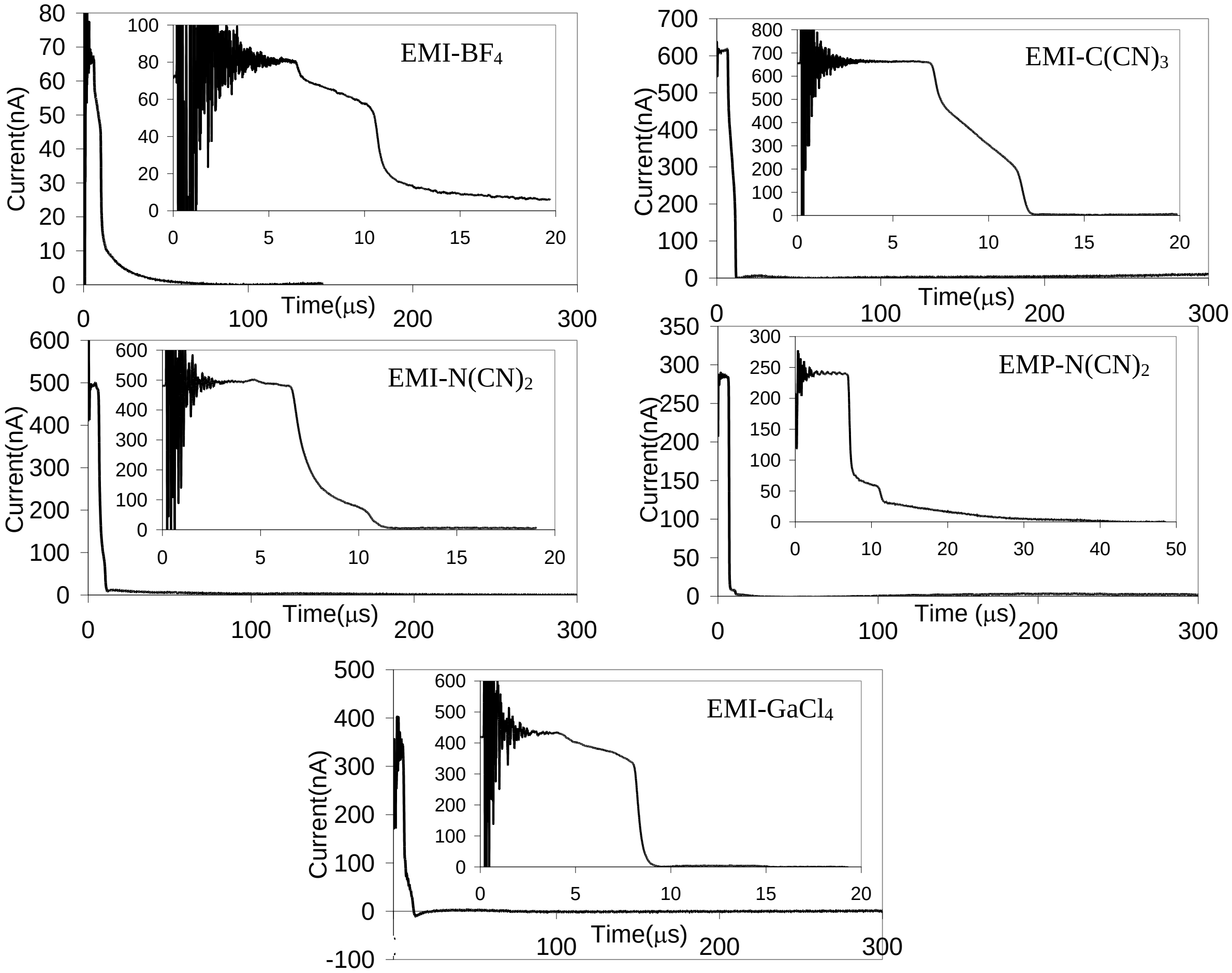

Figure 7: TOF spectra comparing the emissions from EMI-$BF_4$ with those of four other ionic liquids of relatively high surface tension and electrical conductivity.

[37]Y. Yoshida, O. Baba, and G. Saito, J. Phys. Chem. B 111 (18), 4742-4749 (2007).

[38]D. R. MacFarlane, J. Golding, S. Forsyth, M. Forsyth, and G. B. Deacon, Chem. Commun. 1430 (2001).